\documentclass[twocolumn,showpacs,amsmath,amssymb,pra]{revtex4}

\usepackage{amsfonts}

\usepackage{graphicx}
\usepackage{dcolumn}
\usepackage{bm}
\usepackage{epsfig}
\usepackage{amsmath}
\usepackage{mathrsfs}
\usepackage{CJK}

\newcommand{\be}{\begin{equation}}
\newcommand{\ee}{\end{equation}}
\newcommand{\bey}{\begin{eqnarray}}
\newcommand{\eey}{\end{eqnarray}}
\newcommand{\ba}{\begin{array}}

\newcommand{\ea}{\end{array}}
\newcommand{\bi}{\begin{itemize}}
\newcommand{\ei}{\end{itemize}}
\newcommand{\bem}{\begin{enumerate}}
\newcommand{\eem}{\end{enumerate}}
\newcommand{\bw}{\begin{widetext}}
\newcommand{\ew}{\end{widetext}}

\newcommand{\ra}{\rangle}
\newcommand{\la}{\langle}

\newcommand{\ww}{\widetilde}
\newcommand{\bp}{{\bf p}}

\newcommand{\bx}{{\bf x}}

\newcommand{\E}{{\cal E}}

\renewcommand{\vec}[1]{\bm{#1}}

\begin{document}

\title{Signature of existence of a BEC-type state in a dilute gas above the BEC transition temperature
}

 \author{Yinbiao Yang and Wen-ge Wang\footnote{Email address: wgwang@ustc.edu.cn}}

 \affiliation{
 Department of Modern Physics, University of Science and Technology of China,
 Hefei 230026, China
 }

\date{\today}%

\begin{abstract}

  We study quantum coherence properties of a dilute gas at temperatures above,
  but not much above the transition  temperature of Bose-Einstein condensation (BEC).
  In such a gas, a small proportion of the atoms may possess coherence lengths longer than the mean
  neighboring-atomic distance, implying the existence of quantum coherence more than that
  expected for thermal atoms.
  Conjecturing that a part of this proportion of the atoms may lie in a BEC-type state,
  some unexplained experimental results [Phys.Rev.A, \textbf{71}, 043615 (2005)] can be explained.
\end{abstract}





\maketitle

\section{Introduction}\label{sect-intro}

 Quantum coherence at the macroscopic and mesoscopic scales is a topic of interest in a variety of fields.
 A well-known example is the Bose-Einstein condensation (BEC) formed in a dilute gas of identical atoms,
 when the temperature is dropped below a
 transition temperature $T_c$ \cite{Anderson,Davis,Bradley,Dalfovo,Huang,Leggett,Pethick}.
 One interesting question is, at temperatures above and within the same
 order of magnitude of $T_c$, whether the atoms may possess quantum coherence more than
 that expected for thermal atoms.
 In fact, at temperatures a little above $T_c$,
 the stochastic-Gross-Pitaevskii-equation approach predicts that
 a fraction of the atoms may lie in a BEC-type state \cite{Gardiner08}.
 Some other approaches give more or less similar predictions, restricted to the region
 around $T_c$ \cite{Stoof99,Stoof01,Gardiner03,Berloff,Blakie,Bezett,Calzetta,QK,You}.
 However, for temperatures several-times higher than $T_c$,
 these theoretical approaches do not give a clear prediction for the above-mentioned quantum coherence.

 Interestingly, on the other hand, experimental evidence exists for the above-discussed quantum coherence.
 It is given by a recent experiment, in which contrasts of the interference patterns formed by
 a dilute gas of atoms were measured \cite{Miller}.
 It was observed that, at some temperatures above $T_c$ and under some values of controlling
 parameters, the measured contrasts
 are obviously higher than those predicted for thermal atoms.
 This implies the existence of quantum coherence more than that expected for thermal atoms.
 Theoretical explanation to this extra coherence is still absent
 and a purpose of this paper is to take a first step to it.

 To explain the approach we are to take, let us first consider
 a gas of thermal (identical) atoms at a temperature much higher than $T_c$.
 At such a high temperature, the indistinguishability of the atoms usually do not have
 a significant effect and, as a result, the atoms can be treated effectively as distinguishable particles.
 As an approximation, a single atom in the gas can be treated as a
 quantum Brownian particle interacting with a thermal bath.
 Below, we call this approximation the distinguishable-particle approximation.
 Behaviors of a quantum Brownian particle have been studied extensively in passed years
 (see, e.g., Refs.\cite{Caldeira,Hu,Halliwell,Ford,Eisert,Hornberger,Weiss} and references therein).
 Due to environment-induced decoherence \cite{Zeh70,Zurek81,Zurek03,JZKGKS03,Schloss04},
 the reduced state of the particle
 may approach an approximately-diagonal form in a basis given by Gaussian wave packets \cite{Ford,Eisert},
 which is usually referred to as preferred (pointer) basis
 \cite{Zurek81,Zurek03,rpl-12,Schloss04}.
 This implies that the atom can be effectively described by a mixture of Gaussian packets.

 The width of a Gaussian wave packet discussed-above gives a measure to the coherence length of the
 related atom.
 It can not keep constant due to wave-packet expansion \cite{Zurek05}, in other words,
 it should be distributed over some finite region.
 An intriguing question is what may happen to those atoms whose coherence lengths
 are of the scale of the mean neighboring-atomic distance.
 The above-discussed distinguishable-particle approximation fails for these atoms,
 since the related Gaussian packets usually have non-negligible overlap with
 those of the neighboring atoms.
 It would be reasonable to expect that these atoms, at least some of them, may possess coherence more than
 that expected for thermal atoms.

 At high temperatures, the proportion of the above-discussed atoms
 with long coherence lengths should be too small to induce any notable effect in most cases.
 However, when the temperature drops to the order of $T_c$, attention should be paid to them.
 In fact, as well known, when the temperature becomes close to $T_c$, the thermal de Broglie
 wavelength reaches the scale of the mean neighboring-atomic distance \cite{Pethick}.
 In addition, the mean coherence length of the atoms, given by the mean width of the related
 Gaussian wave packets, is of the order of the thermal de Broglie wavelength \cite{Eisert}.
 These two points suggest an intuitive picture for a gas to enter into a BEC state,
 that is, loosely speaking, it may happen when the mean coherence length obtained under the
 distinguishable-particle approximation reaches the scale of the mean neighboring-atomic distance.
 Based on this picture, it would be reasonable to assume that (part of) the atoms with sufficiently long
 coherence lengths may lie in some BEC-type state.

 In this paper, we show that, making use of the above-discussed assumption about
 the existence of some BEC-type state for some atoms,
 a semi-quantitative explanation can be gained to main features of the experimentally-observed,
 unexpectedly-high contrasts discussed above.
 Specifically, the paper is organized as follows.
 In Sec.\ref{sect-ana}, we discuss two models for a gas at temperatures above $T_c$:
 a simple thermal model, in which the atoms are treated as thermal atoms, and a hybrid model,
 in which most of atoms are thermal atoms while a small proportion of the atoms lie in a BEC-type state.
 Based on decoherence arguments, we derive an expression for the temperature-dependence of the
 proportion of the atoms with long coherence lengths.
 In Sec.\ref{sect-expm}, we discuss predictions of the hybrid model for the contrasts of the interference
 patterns observed in the experiments in Ref.\cite{Miller}
 and compare the predictions with experimental results.
 Finally, concluding remarks and discussions are given in Sec.\ref{sect-conclusion}.

 \section{Two models for a gas of atoms at temperatures above $T_c$}
 \label{sect-ana}

 In this section, we discuss two models for a gas of $N$ identical atoms:
 a simple thermal model for high temperatures in Sec.\ref{sect-simple-model}
 and a hybrid model for temperatures above and of the order of the BEC-transition temperature $T_c$
 in Sec.\ref{sect-hybrid-model}.
 The temperature-dependence of the small proportion of the atoms with long coherence lengths
 is discussed in Sec.\ref{sect-Pcon}.

 \subsection{A simple thermal model for high temperatures}
 \label{sect-simple-model}

 We neglect the internal motion of the atoms.
 At high temperatures with $T\gg T_c$, as discussed in Sec.\ref{sect-intro},
 one may take the distinguishable-particle approximation.
 At the end of this section, we discuss that this approximation leads to a self-consistent
 picture for the motion of most of the atoms.

 We use $S$ to denote a considered, central atom and use $\E$ to denote the rest of the atoms,
 as the environment of $S$.
 The state of the total system is denoted by $|\Psi\ra$.
 Assuming that the environment can be regarded as a heat bath,
 the central atom behaves like a particle undergoing a quantum Brownian motion, which
 has been extensively studied \cite{Caldeira,Hu,Halliwell,Ford,Eisert,Hornberger,Weiss}.
 Coherent states are known to give approximate preferred states,
 in which the reduced density matrix (RDM) of the central particle,
 $\rho^{re}= {\rm Tr}_\E(|\Psi\ra\la \Psi|)$, approaches an
 approximate diagonal form beyond some finite time scale \cite{Eisert},
\begin{equation}\label{rho-diag}
 \rho^{re}(t) \approx 
  \int \mathrm{d}\mu_\alpha \rho_\alpha(t) |\alpha,\xi_0\ra \la \alpha,\xi_0|.
\end{equation}
 Here, $|\alpha,\xi_0\ra $ indicates a coherent state, which is a Gaussian wave packet in the case
 studied here,
 centered at $\alpha = (\bx,\bp)$ in the phase space, with a fixed dispersion $\xi_0$ in the coordinate space.
 This form of the RDM suggests that, effectively, the central atom may be described
 by a mixture of Gaussian wave packets, with probabilities $\rho_\alpha$.
 If further assuming that $\rho_\alpha(t)$ has reached a stationary solution with a Boltzmann form,
 one gets a simple thermal model for the atoms in an equilibrium state,
 which has been used in Ref.\cite{Miller}.

 In Eq.(\ref{rho-diag}), the fixed dispersion $\xi_0$ is temperature-dependent \cite{Eisert},
\begin{equation}\label{xi0}
 \xi_0 =\frac{\hbar}{\sqrt{2mk_B T}} = \frac{1}{2 \sqrt{\pi}} \lambda_T,
\end{equation}
 where $\lambda_T$ is the thermal de Broglie wavelength.
 Here, $\hbar$ and $k_B$ denote the Planck constant and the Boltzmann constant, respectively.
 As pointed in Ref.\cite{Zurek05}, due to wave packet expansion, the dispersion of the packet in fact
 can not keep constant.
 The above-discussed fixed-value $\xi_0$ can be regarded as corresponding to the mean value of the dispersion.
 At $T\gg T_c$, $\xi_0$ is much smaller than the mean distance between neighboring atoms,
 denoted by $d_a$ in what follows.
 As a result, most of the Gaussian packets of neighboring atoms have negligible overlap
 in the coordinate space.
 This justifies validity of the distinguishable-particle approximation.

 \subsection{A hybrid model for $T$ of the order of $T_c$} \label{sect-hybrid-model}

 When the temperature $T$ drops to the order of $T_c$, the dispersion of the packets discussed above
 reaches the order of $d_a$ on average, hence, its variation can no longer be neglected.
 To study properties of the central atom in this case,
 we expand the state vector of the total system in the following form, with the dispersion
 as a variable, denoted by $\xi$,
 \begin{equation} \label{Psi-ax}
 |\Psi (t) {\rangle} = \int d\mu_\alpha d\xi |\alpha\xi\ra |\Phi_{\alpha\xi}^{\E}(t)\ra,
 \end{equation}
 where $|\Phi_{\alpha\xi}^{\E}(t)\ra$ are the corresponding components of the environment.
 We assume that decoherence has happened, such that Gaussian wave packets are approximately
 preferred states \cite{Zurek05} and the components $|\Phi_{\alpha\xi}^{\E}(t)\ra$ of the environment
 satisfy
\begin{equation}\label{decoh-g}
 \la \Phi_{\alpha'\xi'}^{\E} (t)| \Phi_{\alpha\xi}^{\E}(t) \ra \approx 0 \
\end{equation}
 for $\alpha$ not close to $\alpha'$ and for $\xi$ not close to $\xi'$.
 Then, the RDM has approximately the following `diagonal' form,
\begin{equation}\label{rho-ax}
 \rho^{re}(t) \approx \int d\mu_\alpha d\xi \rho_{\alpha \xi}(t) |\alpha \xi \ra \la \alpha \xi |,
\end{equation}
 where
\begin{equation}\label{rho-ax-P}
 \rho_{\alpha \xi}(t) = \la \Phi_{\alpha\xi}^{\E} (t)| \Phi_{\alpha\xi}^{\E}(t) \ra .
\end{equation}
 According to Eq.(\ref{rho-ax}), effectively, the atom can be regarded as lying
 in a mixed state, i.e., in a mixture of $|\alpha \xi\ra $ with probabilities $\rho_{\alpha \xi}(t)$.

 In the mixed-state description discussed above, those Gaussian wave packets
 $|\alpha\xi\ra$ with large dispersions, $\xi \gtrsim d_a$, may induce a problem.
 That is, they usually have non-negligible overlap with Gaussian wave packets of the neighboring atoms,
 as a result, symmetrization of the whole wave function does not allow to treat the corresponding
 atoms as distinguishable particles.
 This is in confliction with the distinguishable-particle approximation, which is the starting point of
 the above approach.
 This confliction suggests that the related atoms may possess quantum coherence more
 than that expected for uncorrelated thermal atoms.

 Then, what type of quantum coherence may the corresponding atoms have?
 As discussed in the section of introduction, for temperatures a little above $T_c$, some fraction
 of the atoms may lie in a BEC state.
 It would be natural to expect that, at temperatures not so close to $T_c$ but still of the
 order of $T_c$, a small proportion of the atoms may still lie in some BEC-type state.

 To be specific, we recall that, as discussed previously, loosely speaking, BEC transition happens
 when the mean coherence length of the atoms obtained in the
 distinguishable-particle approximation, which is of the order of the thermal de Broglie wavelength,
 reaches the order of the mean neighboring-atomic distance $d_a$.
 Based on this understanding of BEC transition,
 we make the following conjecture, which is the basic assumption of this paper.
 That is, approximately, a BEC-type state may develop among those atoms connected by
 the relation of mutual-coherence.
 Here, two atoms are said to have a \emph{mutual-coherence} relation,
 if they are associated with two Gaussian wave packets whose
 coherence lengths are longer than the distance between the centers of the two packets.
 Two atoms in mutual-coherence with a same third atom are regarded as being in mutual-coherence, too.

 To be quantitative, the probability for an atom to have a coherence length characterized by $\xi$ is
 written as
\begin{equation}\label{P-xi}
 P(\xi,t) = \int d\mu_\alpha \rho_{\alpha \xi}(t).
\end{equation}
 We use $P_{\rm con}$ to denote the probability for an atom to lie in a BEC-type state.
 The above conjecture implies that this probability should be proportional
 to the probability for the Gaussian wave packets to have dispersions of the order of
 or larger than $d_a$.
 Hence, it can be written as
\begin{equation}\label{P-con} 
 P_{\rm con}(t) = c_1\int^{\infty}_{d_c}  P(\xi,t) {d}\xi ,
\end{equation}
 where, $d_c$ is of the order of $d_a$.
 In what follows, for simplicity in discussion, we assume that $d_c=d_a$,
 since generalization of the results to be given below to the case of $d_c$ not equal to $d_a$
 is straightforward.

 In Eq.(\ref{P-con}), we introduce a parameter $c_1$, since according to the above conjecture
 only atoms connected by the relation of mutual-coherence may lie in the same BEC-type state.
 Obviously, $c_1 \le 1$.
 At temperatures $T$ not high, there exists only one big body of the atoms that are connected by
 the mutual-coherence relation.
 In this case, $c_1$ is approximately temperature-independent.
 In what follows, we discuss this case \cite{footnote-c1}.

 For $c_1 < 1$, there exist atoms associated with Gaussian wave packets possessing large dispersions,
 which do not contribute to the BEC-type state.
 Since the population of these atoms is small and the coherence among them is not as strong as that in
 the BEC-type state, it would be reasonable to expect that these atoms should give small contribution to
 the contrast of the interference pattern formed by the gas.
 Hence, to simplify the discussion, as an approximation, we assume that
 these atoms can be treated as thermal atoms.

 Finally, we get the following hybrid model for a gas in an equilibrium state
 with a temperature $T$ above and of the order of $T_c$.
 That is, part of the atoms associated with $|\alpha\xi\ra$ of $\xi \gtrsim d_a$
 lie in a BEC-type state,
 while, other atoms are thermal atoms described by the simple thermal model discussed
 in the previous section.
 Here, to be specific, some atoms lying in a BEC-type state means that they can be described by a same
 single-particle wave function when computing the interference pattern they generate.

 \subsection{Temperature-dependence of $P_{\rm con}$} \label{sect-Pcon}

 In this section, we discuss dependence of $P_{\rm con}$ on the temperature $T$.
 Let us first discuss the time variation of $P(\xi,t)$ in Eq.(\ref{P-xi}).
 It is mainly determined by competition of the following two aspects
 of the Schr\"{o}dinger evolution of $|\alpha\xi\ra|\Phi_{\alpha\xi}^{\E}(t)\ra$
 on the right hand side of Eq.(\ref{Psi-ax}).
 On one hand, expansion of the wave packet $|\alpha\xi\ra$ converts
 $\rho_{\alpha\xi}$ in Eq.(\ref{rho-ax-P}) to $\rho_{\alpha'\xi' }$ with a larger dispersion $\xi'>\xi$.
 On the other hand, the interaction between the central atom and other atoms may induce decoherence,
 changing $|\alpha\xi\ra |\Phi_{\alpha\xi}^{\E}\ra$
 to a superposition of $|\alpha'\xi'\ra |\Phi_{\alpha'\xi'}^{\E}\ra$ with smaller dispersions $\xi'<\xi$
 and with almost orthogonal $|\Phi_{\alpha'\xi'}^{\E}\ra$.
 This decoherence process converts $\rho_{\alpha\xi}$ to $\rho_{\alpha'\xi' }$ with
 smaller dispersion $\xi'<\xi$.
 Therefore, the probability $P(\xi,t)$ at a time $t$ has two sources:
 The first is due to wave-packet expansion
 from $|\alpha'\xi'\ra|\Phi_{\alpha'\xi'}^{\E}(t')\ra$ with $\xi'<\xi$ at some previous time $t'$,
 and the second is due to decoherence
 from $|\alpha''\xi''\ra|\Phi_{\alpha''\xi''}^{\E}(t'')\ra$ with $\xi''>\xi$ at some previous time $t''$.

 In Eq.(\ref{P-con}), only those $P(\xi,t)$ with $\xi \ge d_a$ contribute to $P_{\rm con}$.
 At temperatures $T$ obviously higher than (still of the order of) $T_c$,
 $d_a$ is obviously larger than the mean value of $\xi$.
 Physically, one can assume that $P(\xi,t)$ decreases sufficiently fast with increasing
 $\xi$ beyond $\xi=d_a$.
 (Later we show that this assumption leads to a self-consistent result.)
 Then, for $P(\xi,t)$ with $\xi \ge d_a$,
 the above-discussed contribution from the second source
 is small, compared with that from the first one, and can be neglected.

 Then, for large $\xi$, we get the following expression of $P(\xi,t)$, in terms of $P(\xi',t')$
 with $\xi' <\xi$ and $t'<t$,
 \begin{equation}\label{Delta-P}
 P(\xi,t) \approx P(\xi',t') - \eta_{_T}(\xi') P(\xi',t') \Delta t ,
 \end{equation}
 where $\Delta t=t-t'$, the second term on the right hand side represents the effect of decoherence
 which converts $|\alpha'\xi'\ra |\Phi_{\alpha'\xi'}^{\E}\ra$ to superpositions of narrower
 wave packets of the central atom, and $\eta_{_T}(\xi')$ indicates the rate of this decoherence process.
 The two variables $\xi$ and $\xi'$ are connected by the relation $\xi=\xi'+v_e \Delta t$,
 where $v_e$ is the expanding speed of the packet.
 The speed $v_e$ is determined by the width of the initial packet whose expansion gives contributions to
 both $P(\xi,t)$ and $P(\xi',t')$, hence, $v_e$ is $\xi$-independent.
 Since as discussed above the dispersion $\xi$ has a mean value given by $\xi_0$ in Eq.(\ref{xi0}),
 in most cases $v_e$ is approximately determined by $\xi_0$, hence, it is temperature-dependent.
 According to standard textbooks [see Eq.(\ref{sp}) to be cited below],
 $v_e \propto 1/\xi_0$, as a result, $v_e\propto T^{1/2}$.
 Therefore, we write $v_e=u_0 T^{1/2}$ with $u_0$ approximately temperature-independent.

 In an equilibrium state, the probability $P(\xi,t)$ is time-independent, denoted by $P(\xi)$.
 For large $\xi$, Eq.(\ref{Delta-P}) shows that this distribution satisfies
\begin{equation}\label{dP-dxi}
 \frac{\mathrm{d} P(\xi)}{\mathrm{d} \xi} \approx -\frac{\eta_{_T}(\xi)}{u_0T^{1/2}} P(\xi),
\end{equation}
 where $v_e=u_0 T^{1/2}$ has been used.
 Equation (\ref{dP-dxi}) has a solution,
   \begin{equation}\label{P-xi-int} 
   P(\xi) \approx a_{0} \exp \left \{ -\frac{1}{u_0T^{1/2}}\int \mathrm{d}\xi \eta_{_T}(\xi)\right \},
   \end{equation}
 where $a_0$ is an integration constant.

 We assume that the decoherence rate $\eta_{_T}(\xi)$ has the following dependence on $\xi$ and $T$,
\begin{equation}\label{eta-T}
 \eta_{_T}(\xi)  \simeq a_1 \xi^\gamma T^\beta,
\end{equation}
 with a parameter $a_1$ independent of $\xi$ and $T$.
 To get the values of $\gamma$ and $\beta$, we note that
 the decoherence is induced by collisions among the atoms.
 This implies that, approximately,
 $\eta_{_T}$ should be proportional to the number of collisions per time unit.
 First, it should be proportional to the mean speed of the atoms,
 hence, $\eta_{_T} \sim \sqrt T$, giving $\beta = 1/2$.
 Second, since the collision number is approximately proportional to the cross sections
 of the Gaussian packets, one has $\eta_{_T} \sim\xi^2$, giving $\gamma=2$.
 For $\beta = 1/2$ and $\gamma = 2$, Eq.(\ref{P-xi-int}) gives
\begin{equation}\label{P-xi-2} 
   P(\xi) \approx a_{0} \exp \left \{ -\frac{a_1 }{3u_0} \xi^{3} \right \}.
\end{equation}
 Equation (\ref{P-xi-2}) shows that our previous assumption about the fast decay of $P(\xi,t)$
 for large $\xi$ is self-consistent.

 As to the parameter $a_0$, we note that
 due to the $T$-dependence of the decoherence rate $\eta_{_T}(\xi)$ with $\beta =1/2$,
 the variation rate of $P(\xi)$ in Eq.(\ref{dP-dxi}) with respect to $\xi$
 is in fact $T$-independent.
 This suggests that the parameter $a_0$ may be $T$-independent.
 Below, we assume that $a_0$ is either $T$-independent or changes slowly with $T$.

\begin{figure} 
 \includegraphics*[width=\columnwidth]{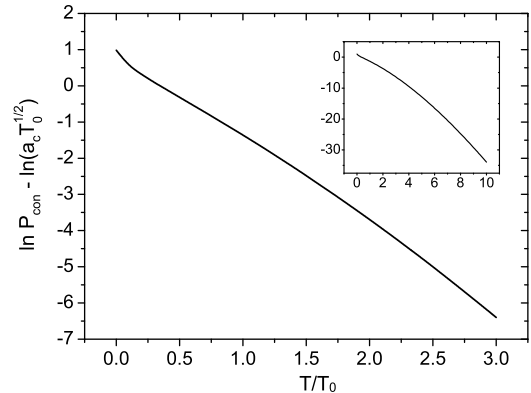}
 \hspace{2.2cm}\caption {
 Variation of $P_{\rm con}/(a_c T_0^{1/2})$, in the logarithm scale,
 with $T/T_0$ for $\gamma=2$ and $\beta=0.5$ [see Eq.(\ref{P_con})].
 It has approximately an exponential decay for ${T}$ below $3 T_0$.
 Inset: The decay is even faster for larger $T/T_0$.
 }
\label{decay}
  \end{figure}

 Now, we compute the proportion $P_{\rm con}$.
 Substituting Eq.(\ref{P-xi-int}) with Eq.(\ref{eta-T}) into Eq.(\ref{P-con}), direct derivation
 shows that, in terms of $z=(\xi/d_a)^{\gamma+1}$, $P_{\rm con}$ can be written as
   \begin{equation}
   P_{\mathrm{con}} \approx \frac{d_a a_0 c_1}{\gamma+1}\int^{\infty}_{1} \mathrm{d} z
   z^{-\frac{\gamma}{\gamma+1}}\exp\left(-\frac{a_1 d_a^{\gamma+1}}{(\gamma+1)u_0}
   T^{\beta-1/2} z \right).
   \end{equation}
 In the experiment we are to discuss, the mean atomic distance $d_a$ is proportional to $\sqrt T$,
 therefore, we write
\begin{equation}\label{da}
 d_a=a_2 T^{1/2}
\end{equation}
 [see Eq.(\ref{da-app}) to be given later for an explicit expression of $d_a$].
 Then, noticing that $u_0, a_1$, and $ a_2$ are $\xi$-independent, we have
   \begin{equation}
   P_{\mathrm{con}} \approx a_c T^{1/2} E_{\frac{\gamma}{\gamma+1}}
   \left(\left(\frac{T}{T_0}\right)^{\gamma/2+\beta}\right), \label{P_con}
   \end{equation}
 where $E_n(x)$ is a function defined by
\begin{equation}
 E_n(x)=\int_1^{\infty} z^{-n} e^{-x z}dz,
\end{equation}
 and
\begin{eqnarray}\label{T0}
T_0&=&\left[\frac{a_1 a_2^{\gamma+1}}{(\gamma+1)u_0}\right]^{-\frac{1}{\gamma/2+ \beta}},
\\ a_c&=&\frac{a_0 a_2 c_1}{\gamma+1}. \label{a-c}
\end{eqnarray}
 The proportion $P_{\rm con}$ has an exponential-type decay in the temperature region
 of interest here (see Fig.\ref{decay}).

\section{Experimental evidence of the existence of a BEC-type state above $T_c$}
\label{sect-expm}

 In this section, we show that some of the unexplained experimental results given
 in Ref.\cite{Miller} can be explained in the hybrid model introduced above.

 \subsection{Experimental results of Ref.\cite{Miller}}\label{sect-experiment}

 In this subsection, we summarize experimental results given in Ref.\cite{Miller},
 which are of relevance to the study of this paper.
 In an experiment discussed there, the contrast of the interference pattern formed by a
 cloud of $N$ atoms with mass $m$ is measured.
 Initially, the cloud, confined by a harmonic trap with a frequency $\omega$, is prepared in a thermal state
 at a temperature $T$, which is above the BEC transition temperature $T_c$.
 Shortly after being released from the trap, the gas is exposed to two Bragg beams successively.
 Each Bragg beam has a shining period $\tau_p$.
 This process creates an identical copy of the initial cloud, separated by a distance denoted by $d$.
 Then, the cloud and its copy expand freely and, after a period of flying time $\tau_f=48 {\rm ms}$,
 form an interference pattern with a contrast denoted by $C_{\rm ex}$.

 To analyze the experimental results, the simple thermal model discussed in Sec.\ref{sect-simple-model}
 was studied in Ref.\cite{Miller}, with dispersion of the Gaussian packets given by
 the thermal de Broglie wave length, $\lambda_T{=}{h}/{\sqrt{2\pi m k_B T}}$.
 This model predicts the following contrast for the interference pattern,
\begin{equation}\label{Cth}
 C_{\mathrm{th}}
 =\exp \left( -\frac{2\pi^2R_T^2}{\lambda_f^2}\right),
\end{equation}
 with the subscript `th' standing for thermal, where
\begin{eqnarray}\label{lf}
\lambda_f=\frac{h\tau_f}{md}, \quad R_T=\sqrt{\frac{k_B T}{m \omega^2}}.
\end{eqnarray}
 The quantity $\lambda_f$ gives the fringe spacing of the interference pattern.

 The following results were reported in Ref.\cite{Miller}, concerning the contrast $C_{\rm ex}$.
\begin{enumerate}
  \item[(i)] There exists approximately a temperature, which we denote by $T_d$, below which
  $C_{\rm ex}$ are close to $C_{\mathrm{th}}$ and above which
  $C_{\rm ex}$ are higher than $C_{\rm th}$.
  \item[(ii)] The temperature $T_d$ changes notably with the shining length $\tau_p$ of the Bragg beams,
  but, is not so sensitive to the distance $d$.
  \item[(iii)] The Bragg beams are velocity-selective for relatively long period $\tau_p$.
\end{enumerate}
 More specifically, for the point (ii), with $T_c \approx 0.6 \mu K$, in the case of $\tau_p= 10 \mu s$,
 $T_d \approx 3 \mathrm{\mu K}$ for $\lambda_f=340\mathrm{\mu m}$,
 and $T_d \approx 2\mathrm{\mu K}$ for both $\lambda_f=230 \mathrm{\mu m}$ and
 $\lambda_f=170\mathrm{\mu m}$.
 In the case of $\tau_p=30\mu s$,  $T_d \approx 1 \mathrm{\mu K}$ for $\lambda_f=170 \mathrm{\mu m}$.
 Related to the point (iii), within the simple thermal model,
 the velocity-selection effect can not explain the observed, unexpectedly high contrast of
 $C_{\rm ex}$ at $T>T_d$ \cite{Miller}.

 \subsection{Detailed predictions of the simple thermal model} \label{thermal-model}

 Before discussing predictions of the hybrid model and comparing them with
 the above-discussed experimental results, it would be
 useful to discuss in more detail predictions of the simple thermal model.
 In this model, when the cloud is released from the trap at an initial time $t=0$,
 the atoms are described by an ensemble-mixture of Gaussian wave packets, with wave functions
 $\varphi(\bx,\alpha_0)$,
\begin{equation}
 \varphi(\bx ,\alpha_0) = A^3_{\xi}
 \exp\left[\frac{i\vec{p}_0\cdot(\vec{x}-\vec{x}_0)}{\hbar}-\frac{(\vec{x}-\vec{x}_0)^2}{4\xi^2}\right],
 \label{MUS}
\end{equation}
 where $\alpha_0 = (\vec{x}_0,\vec{p}_0)$ indicate centers of the packets and
\begin{equation}\label{A-xi}
 A_{\xi} = {(2\pi)^{-1/4}\xi^{-1/2}}
\end{equation}
 is the normalization coefficient.
 We consider a constant dispersion $\xi$,
 of the order of the thermal de Broglie wave length, $\xi \sim \lambda_T$.
 The values of $\alpha_0 $ are assumed to obey the Boltzmann distribution,
\begin{equation}\label{f-dis}
  f(\vec{x}_0,\vec{p}_0) =\left(\frac{\hbar \omega}{k_B T}\right)^3 \exp \left[ -\frac{1}{k_{B}T}
  \left( \frac {\vec{p}_0^2}{2m}+\frac{m\omega^2\vec{x}_0^2}{2} \right) \right] ,
\end{equation}
 where $m\omega^2\vec{x}_0^2/2$ is the potential generated by the harmonic trap,
 centered at the origin of the coordinate space.
 Direct computation shows that, under this distribution,
 the standard deviation of the position of an atom in each direction is given by
 $R_T$ in Eq.(\ref{lf}).
 Thus, the majority of the atoms lies within a sphere with a radius $R_T$.
 Using this property, we get the following estimate to the mean neighboring-atomic distance
 in the initial cloud,
\begin{equation} \label{da-app}
  d_a  \approx R_T \left( \frac{4\pi}{3N} \right)^{1/3} .
\end{equation}
 Hence, the parameter $a_2$ in Eq.(\ref{da}) has approximately the expression
 $a_2 \approx \sqrt{k_B/m\omega^2} \sqrt[3]{4\pi/3N}$.

 Let us use $t_b$ to denote a time immediately beyond the second Bragg beam.
 We assume that $t_b$ is short, such that the wave packets still have Gaussian shape.
 If a Bragg beam converts a packet into two packets, the two Bragg beams
 convert an initial Gaussian wave packet into four packets.
 Two in the four packets have the same mean velocity and we study the interference pattern formed by them.
 We use $\vec{d}$ to indicate the displacement of the two packets at this time, with $|\vec{d}|=d$,
 and take its direction as the $x$-direction of the coordinate system.
 The $z$-direction is taken to be perpendicular to the plane of the measured interference pattern.

 The above-discussed two packets are written as
\begin{equation}
 \psi(\vec{x},\alpha_0,t_b) =J[ \psi_1(t_b) + \psi_2(t_b)],\label{initial state}
\end{equation}
where
\begin{eqnarray}
 \psi_1(t_b)\rangle \simeq \varphi(\vec{x},\alpha_1), \quad \psi_2(t_b)\simeq \varphi(\vec{x},\alpha_2),
\end{eqnarray}
 with $\alpha_1 = (\vec{x}_0{+}\frac{\vec{d}}{2},\vec{p}_0)$ and
 $\alpha_2 = (\vec{x}_0{-}\frac{\vec{d}}{2},\vec{p}_0)$.
 For brevity, we normalize the vector $\psi(\vec{x},\alpha,t_b)$, with
\begin{equation}
J= \left \{ 2+2\exp \left( -\frac{d^2}{8\xi^2}\right) \cos(\vec{p}_0 \cdot \vec{d}/\hbar) \right \}^{-1/2}.
\end{equation}
 For the parameters used in the experiments and for $\xi \sim \lambda_T$, one has
 $\exp(-\frac{d^2}{8\xi^2}) \ll 1$. This gives $J{\simeq} {1}/{\sqrt 2}$.

 To compute Schr\"{o}dinger evolution of $\psi_1(t)$ and of $\psi_2(t)$, we make use of a result
 given in standard textbooks, namely,
 an initial Gaussian wave packet $\varphi(\vec{x},\alpha_0)$ has the following free expansion\cite{Cohen},
\begin{equation}
\begin{split}
 &A^3_\sigma e^{i\left[\vec{p}_0 \cdot
 \left(\vec{x}-\vec{x}_0\right)\hbar\right]} \\
\times &\exp \left\{ \frac{-\left| \vec{x}-\left(\vec{x}_0
 +\frac{\vec{p}_0t}{m}\right)\right|^2}{4\sigma^2}\left(1-\frac{i \hbar t}{2 m \xi^2}\right)-i\theta(t)
 \right\}, \label{sp}
\end{split}
\end{equation}
 where $A_\sigma$ is defined by Eq.(\ref{A-xi}) ($\xi$ replaced by $\sigma$),
 \begin{eqnarray} \label{sigma}
  \sigma = \sqrt{\left(\frac{\hbar t}{2  m \xi}\right)^2+\xi^2},
\end{eqnarray}
 and $\theta(t)=-\frac{\vec{p}_0^2 t}{2m\hbar}-\frac{3}{2}\arctan\left({\hbar t}/{2 m \xi^2}\right)$.
 Then, it is not difficult to compute
 \begin{equation} \label{psi-t}
 \psi(\vec{x},\alpha_0,t)\simeq \frac 1{\sqrt 2} [\psi_1(t) + \psi_2(t) ]
\end{equation}
 and the density $\rho( \vec{x} ,\alpha_0 ,t) =|\psi(\vec{x},\alpha_0,t)|^2$.
 Integrating $\rho ( \vec{x} ,\alpha_0 ,t)$ thus obtained
 over $\alpha_0 =(\vec{x}_0, \vec{p}_0)$ with the weight
 $f(\vec{x}_0,\vec{p}_0)$, one gets the averaged density, denote by $n(\vec{x},t)$,
 namely, $n= \int \rho f d\alpha_0$.
 For the parameters used in the experiments and for $\xi$ of the order of $ \lambda_T$,
 at the time $t=\tau_f$ at which the interference pattern is measured, one has
\begin{equation}\label{con-xi}
 \xi {\ll} \sqrt{\frac{\hbar \tau_f}{2m}}.
\end{equation}
 Making use of the relation in Eq.(\ref{con-xi}), one can compute
 the density (see Appendix \ref{app-thermal}), obtaining
\begin{eqnarray}  \label{pattern}
&&n(\vec{x},t)
 \simeq   A^6_R e^{-\frac{|\vec{x}|^2}{2 R^2}} \left( 1+C_{\mathrm{th}}\cos\frac{2\pi
x}{\lambda_f} \right), \ \ \label{thermal_interference}
\end{eqnarray}
 where
\begin{equation}
R =\sqrt{R^2_T+\frac{k_B T \tau_f^2}{m}+\sigma^2}= \sqrt{R^2_T(1+\omega^2 \tau_f^2)+ \sigma^2}.
\label{thermal_width}
\end{equation}
 The quantity $R$ gives approximately the size of the expanded thermal cloud.
 Equation (\ref{pattern}) predicts the contrast given in Eq.(\ref{Cth}) for the interference pattern.

 Finally, it would be of interest to give a few words on
 whether notable improvement may be got for the agreement between
 $C_{\mathrm{th}}$ and $C_{\rm ex}$ at temperatures $T>T_d$,
 if the Boltzmann distribution is replaced by the Bose-Einstein distribution.
 We have performed numerical simulations,
 but, have not observed any obvious improvement (see Appendix \ref{sect-Bose}).

 \subsection{Predictions of the hybrid model}\label{the-model}

 In this section, we discuss predictions of the hybrid model introduced in Sec.\ref{sect-hybrid-model}.
 In this model, the density of the cloud is written as
\begin{equation} \label{n}
    n(\vec{x})=n_{\mathrm{th}}+n_{\mathrm{con}},
\end{equation}
 where $n_{\mathrm{con}}$ indicates the contribution from the atoms in the BEC-type state and
 $n_{\mathrm{th}}$ for that from the thermal atoms.
 The density $n_{\mathrm{th}}$ is in fact given by the right hand side of
 Eq.(\ref{thermal_interference}) multiplied by $(1-P_{\rm con})$.

 In the experiments, a contrast was obtained by measuring the intensity of the light reflecting
 from the atoms at the final stage \cite{kt-commu}.
 It corresponds to the contrast given by $n(x,y) = \int n(\vec{x})dz$.
 Since $R {\gg} \lambda_f$, the term $\exp({-\frac{|\vec{x}|^2}{2 R^2}})$ in
 Eq.(\ref{thermal_interference}) can be treated as 1 in the considered region.
 Then, for $n_{\mathrm{th}}(x,y)= \int n_{\mathrm{th}} dz$, we have
\begin{equation}\label{n-th}
n_{\mathrm{th}}(x,y) \simeq (1-P_{\rm con})A^4_R \left(1+ C_{\mathrm{th}}
\cos\frac{2 \pi x}{\lambda_f}\right).
\end{equation}
 We use $L$ to indicate the size of the region occupied by the atoms in the BEC-type state.
 As discussed previously, the atoms in a BEC-type state can be described by a same
 single-particle wave function when computing the interference pattern they generate.
 As long as the absolute value of this wave function changes slowly within the region $L$,
 the density  $n_{\mathrm{con}}(x,y)= \int n_{\mathrm{con}} dz$
 has approximately the following expression,
\begin{equation}\label{n-con}
n_{\mathrm{con}}(x,y) \approx \frac{P_{\rm con}}{L^2}\left( 1+ \cos\frac{2 \pi x}{\lambda_f}\right),
\end{equation}
 independent of the exact shape of the wave function (see Eq.(\ref{un_rho}) in Appendix.\ref{un}).

  \begin{figure}
  \centering
 \includegraphics[width=\columnwidth]{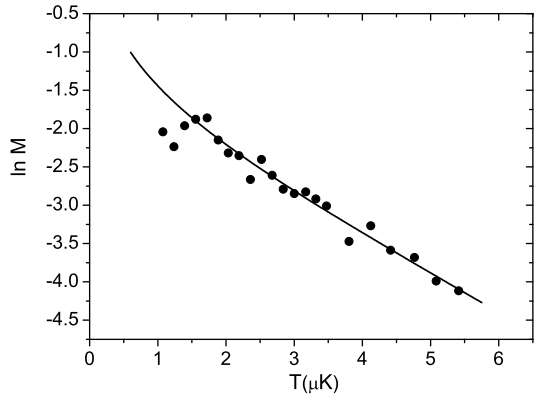}
 \caption { $\ln M$ versus the temperature $T$,
    for $\tau_p{=}30\mathrm{\mu s}$ and $\lambda_f=170 \mathrm{\mu m}$.
    Circles: experimental results, computed from Eq.(\ref{M}) with
    $C$ replaced by the experimentally-obtained contrasts $C_{\rm ex}$ given in Ref.\cite{Miller}.
    Solid curve: analytical predictions computed from Eq.({\ref{fitting}}) with two
  fitting parameters $a{=}0.075 \mathrm{\mu K^{-1.5}}$ and $T_0{=}5.12 \mathrm{\mu K}$.}
  \label{fit}
  \end{figure}

 Making use of Eqs.(\ref{n})-(\ref{n-con}),
 after simple algebra, we get the following expression for the density of the atoms in the $x-y$ plane,
\begin{equation}
n(x,y) \approx G \left( 1+{C} \cos\frac{2\pi x}{\lambda_f}\right),
\end{equation}
 where
 \begin{equation}
 G=(1-P_{\rm con})A^4_R+ \frac{P_{\mathrm{con}}}{L^2}
 \end{equation}
and the modified contrast is given by
\begin{equation}
 {C} = \frac{(1-P_{\rm con})C_{\mathrm{th}}+qP_{\mathrm{con}}}
 {(1-P_{\rm con})+qP_{\mathrm{con}}} \label{C-app}
\end{equation}
 with $q=1/(L^2 {A^4_R })$.
 Making use of the expression of $A_R$ given by Eq.(\ref{A-xi}), one gets
 $ q = \frac {2 \pi R^2}{L^2}$.
 For the parameters used in the experiments, the main contribution to $R$ in
 Eq.(\ref{thermal_width}) is given by the term $\frac{k_B T \tau_f^2}{m}$,
 as a result,
\begin{equation}\label{q}
 \frac qT \approx \frac {2 \pi k_B \tau_f^2}{m L^2}.
\end{equation}
 It is reasonable to assume that the size $L$ of the BEC-type state
 has weak dependence on the temperature $T$.
 Then, the ratio $q/T$ is almost independent of $T$, or changes slowly with $T$.

 The expansion of the BEC-type state should be much slower than that of the thermal cloud.
 This implies that the size $L$ of the BEC-type state should be much
 smaller than the size $R$ of the thermal cloud.
 As a result, $q\gg 1$.
 Hence, even for small $P_{\rm con}$, it is possible for $qP_{\rm con}$ to be not small
 and to give a significant contribution to the predicted contrast $C$ in Eq.(\ref{C-app}).
 But, for sufficiently small $P_{\rm con}$ for which $qP_{\rm con} \ll C_{\rm th}$,
 one has $C\simeq C_{\rm th}$, that is, the prediction of the hybrid model reduces to
 that of the simple thermal model.

 \subsection{Comparison with experimental results}\label{sect-compare}

 \begin{figure}
  \centering
 \includegraphics[width=\columnwidth]{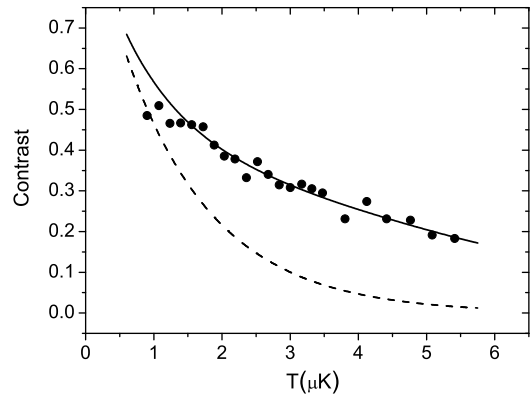}
 \caption{Similar to Fig.\ref{fit}, but for the contrast.
  The theoretical prediction (solid curve) is computed by making use of Eq.(\ref{M})
  with $M$ computed from Eq.({\ref{fitting}}).
  For comparison, $C_{\rm th}$ of the simple thermal model are also plotted (dashed curve).}
  \label{fit1}
  \end{figure}

 In order to compare the above-obtained contrast in Eq.(\ref{C-app})
 and the experimental results, we consider a quantity $M$ defined by
\begin{equation}\label{M}
 M= \frac{C-C_{\mathrm{th}}}{T^{3/2}(1-C)}.
\end{equation}
 Substituting Eq.(\ref{P_con}) with $\gamma =2$ and $\beta =0.5$
 into Eq.(\ref{C-app}), then into Eq.(\ref{M}), and noticing the smallness of $P_{\rm con}$, one gets
 \begin{equation}
  M \approx a E_{\frac{2}{3}}\left(\left(\frac{T}{T_0}\right)^{3/2}\right),
 \label{fitting}
 \end{equation}
 where
 \begin{equation}\label{para-a}
 a=\frac{a_c q}{T} = \frac{a_0 a_2 c_1}{3} \frac{ q}{T}.
 \end{equation}

 Before giving the comparison, we discuss properties of the two parameters $T_0$ and $a$
 in the studied experiments.
 First, let us consider $T_0$ in Eq.(\ref{T0}).
 As discussed in Sec.\ref{sect-Pcon}, the parameters $a_1$ and $u_0$
 are almost temperature-independent.
 The parameter $a_2$ given below Eq.(\ref{da-app}) is also temperature-independent.
 Hence, $T_0$ is temperature-independent.
 The value of $T_0$ is in fact determined by intrinsic properties of the gas
 and by the initial condition (not including the temperature-dependence) of the cloud in the trap.
 Hence, $T_0$ should be also independent of $\tau_p$ and $\lambda_f$.
 Next, for the parameter $a$ in Eq.(\ref{para-a}),
 as discussed in Sec.\ref{sect-Pcon} and Sec.\ref{the-model}, the parameters $a_0$ and $c_1$ and the ratio
 $q/T$ are almost temperature-independent, or change slowly with the temperature $T$.
 Hence, the parameter $a$ is almost temperature-independent, or change slowly with $T$.

 An advantage of considering the quantity $M$, as a function of the temperature $T$,
 is that the parameter $a$ introduces a vertical shift only to $\ln M$.
 Making use of this property and noticing the above-discussed weak dependence of $a$ on $T$,
 it is possible to approximately determine the value of $T_0$
 by a best fitting of the prediction of Eq.(\ref{fitting})
 to the experimental results obtained with one pair of $(\tau_p,\lambda_f)$.
 With the value of $T_0$ thus obtained, one can check whether the prediction of
 Eq.(\ref{fitting}) may be in agreement with
 the experimental results for this pair of $(\tau_p,\lambda_f)$ and, furthermore,
 also check for other pairs of $(\tau_p,\lambda_f)$.

 We use the experimental data obtained with $\tau_p =30 \mathrm{\mu s}$
 and $\lambda_f=170 \mathrm{\mu m}$ to get an estimate to the value of $T_0$.
 We found $T_0 = 5.12 \mathrm{\mu K}$ in a best fitting discussed above.
 As seen in Fig.\ref{fit}, the agreement is good in almost the whole temperature region,
 except for the first two points with $T$ close to $T_d$.
 The agreement is also good in the plot of contrast (Fig.\ref{fit1}).


\begin{figure}
  \centering
 \includegraphics[width=\columnwidth]{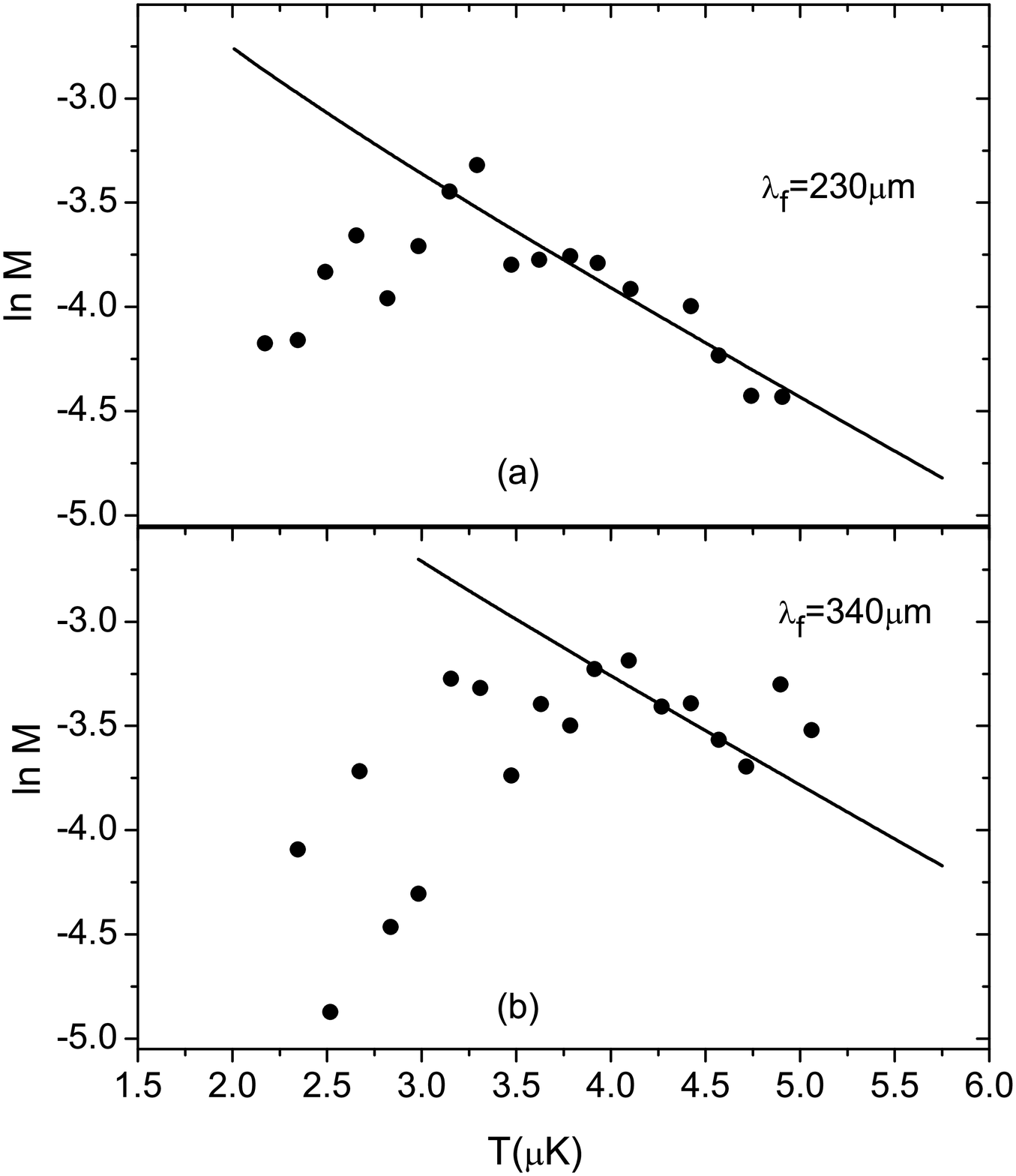}
 \caption{
 Similar to Fig.\ref{fit}, but for $\tau_p{=}10 \mu s$ with a fixed $T_0=5.12 \mu K$.
 The fitting parameter $a=0.044 \mathrm{\mu K^{-1.5}}$ for $\lambda_f=230 \mathrm{\mu m}$ and $a=0.083
 \mathrm{ \mu K^{-1.5}}$ for $\lambda_f=340 \mathrm{\mu m}$.} \label{fit2}
\end{figure}

\begin{figure}
  \centering
 \includegraphics[width=\columnwidth]{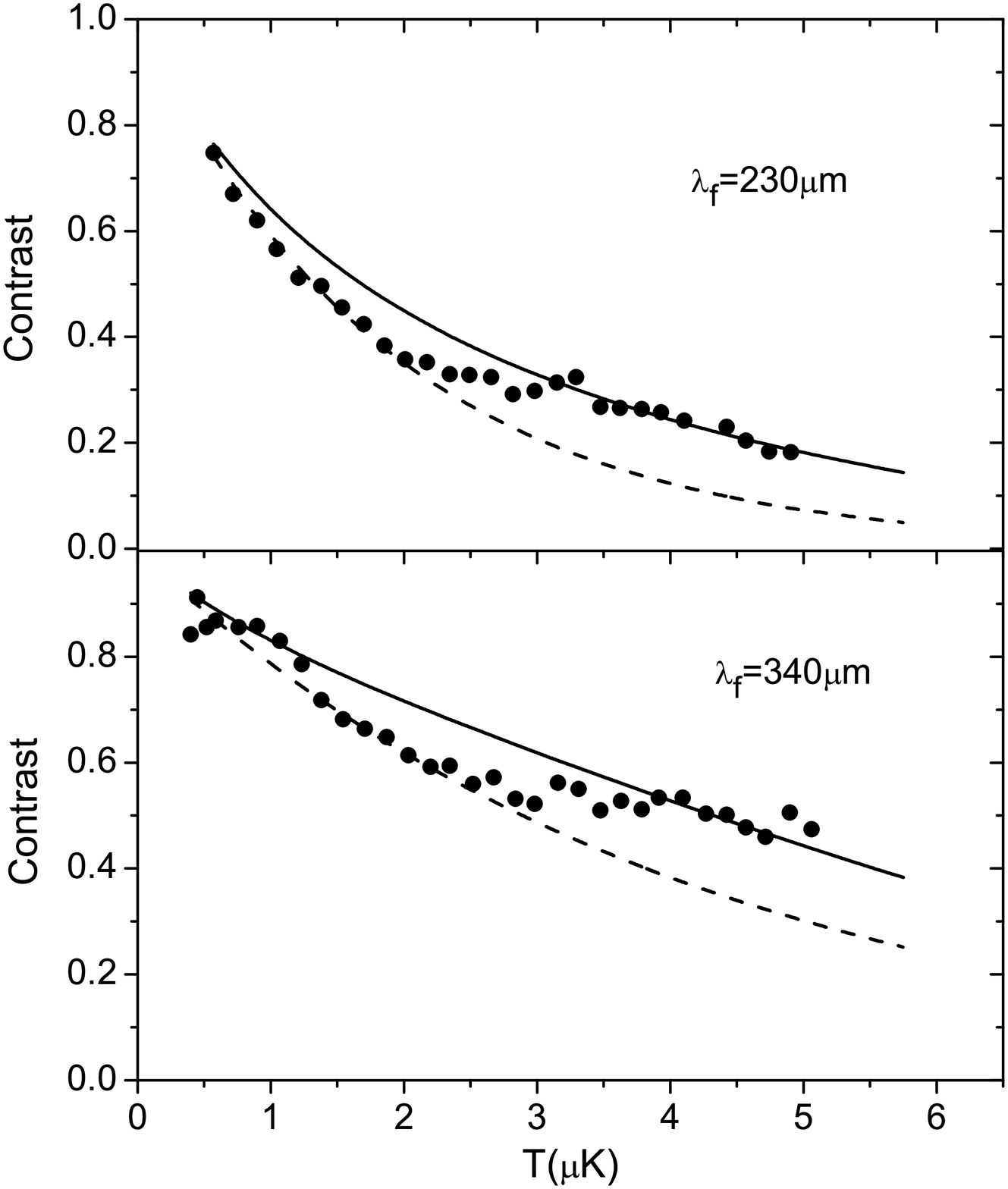}
 \caption{Similar to Fig.\ref{fit2}, but for the contrast. The dashed curve gives $C_{\rm th}$.}
  \label{fit3}
  \end{figure}

 We then study the case of $\tau_p=10 \mathrm{\mu s}$, with the above-obtained
 value of $T_0 = 5.12 \mathrm{\mu K}$.
 We found that, taking the parameter $a$ as a fitting parameter,
 the analytically-predicted $M$ in Eq.(\ref{fitting}) can be in
 agreement with the experimental results in the temperature region of $T \gtrsim T_d+1\mathrm{\mu K}$,
 wherein $C_{\rm ex}$ are obviously larger than $C_{\rm th}$ (see Fig.\ref{fit2}).
 Specifically, for $\lambda_f = 230 \mathrm{\mu m}$ with $T_d\approx 2\mathrm{\mu K}$,
 the agreement is good in the temperature region of $T \gtrsim 3\mathrm{\mu K} $,
 and, for $\lambda_f=340 \mathrm{\mu m}$ with $T_d\approx 3\mathrm{\mu K}$,
 agreement is seen in the region of $T \gtrsim 4\mathrm{\mu K}$ except for the last two points.
 Similar results can also be seen in the plot of contrast (Fig.\ref{fit3}).
 For $\lambda_f=170 \mathrm{\mu m}$, no reasonable comparison can be made, because there is only one or
 two points in the region above $T_d+1\mathrm{\mu K}$.

 On the other hand, for a sufficiently small value of the parameter $a$ such that
 $qP_{\rm con} \ll C_{\rm th}$, the contrasts predicted in the hybrid model in Eq.(\ref{C-app})
 are approximately equal to $C_{\rm th}$ and, hence, close
 to the experimental data $C_{\rm ex}$ in the region of $T<T_d$ (see Fig.\ref{fit3}).
 (The value of $T_0$ does not influence the prediction for $a = 0$.)
 However, for whatever fixed value of the parameter $a$,
 the prediction of Eq.(\ref{C-app}) for the contrast can not be made in agreement with
 the experimental data in the whole temperature region.
 Even we change the value of $T_0$ as well, no obvious improvement has been observed.

 Therefore, in the case of $\tau_p =10 \mathrm{\mu s}$,
 the contrast undergoes a transition approximately in the region $(T_d,T_d+1\mathrm{\mu K})$.
 Below this region the hybrid model works well with $a \approx 0$,
 and above this region the model works well with $a$ as a fitting parameter
 and with $T_0$ fixed at the value determined in the above case of $\tau_p =30 \mathrm{\mu s}$.
 We note that this conclusion can also be regarded as being valid for $\lambda_f=170 \mathrm{\mu m}$.

 To summarize, in the case of $\tau_p =30 \mathrm{\mu s}$,
 the hybrid model can explained main features of the experimental results of the contrasts
 in almost the whole temperature region studied experimentally.
 While, in the case of $\tau_p =10 \mathrm{\mu s}$, in order to
 explained main features of the experimental results by the hybrid model, one needs to assume that
 the contrast undergoes a transition approximately in the region $(T_d,T_d+1\mathrm{\mu K})$,
 below which the parameter $a \approx 0$ and above which $a$ has a nonzero value.

 A hint to a possible origin of the above-discussed transition behavior of the contrast
 lies in an observation made in Ref.\cite{Miller}.
 That is, in the case of $\tau_p =30 \mathrm{\mu s}$
 the two Bragg beams have a significant velocity-selection effect,
 meanwhile, in the case of $\tau_p =10 \mathrm{\mu s}$
 the velocity-selection effect is not so significant.
 This suggests that the velocity-selection effect of the Bragg beams
 may have some relation to the transition behavior of the contrast.

 \section{Concluding remarks and discussions}\label{sect-conclusion}

 In this paper, we study the conjecture that,
 in a gas at a temperature above but not much above the BEC transition temperature,
 there may exist a small proportion of the atoms lying in a BEC-type state.
 We show that experimental support to this conjecture exists,
 that is, in a hybrid model based on this conjecture,
 some unexplained experimental results reported in Ref.\cite{Miller} can be explained.

 What is left unexplained is a transition behavior of the contrast,
 from a relatively-low temperature region, in which the atoms behave like thermal atoms, to a
 relatively-high temperature region,
 in which some of the atoms show quantum coherence like that in some BEC-type state.
 In order to understand this transition behavior, further investigations,
 both experimental and theoretical,  are needed.
 A key point may lie in the role played by a velocity-selection feature of the Bragg beams.

 In the experimental aspect, it would be useful to study possible connection between
 the transition region discussed above and the velocity-selection effect of the Bragg beams,
 for example, in the case of $\tau_p=10 \mathrm{\mu s}$.
 In addition, study of the contrasts at temperatures higher than those reported in Ref.\cite{Miller}
 should be of interest, too.
 In the theoretical aspect, a key problem is whether there may exist any relation
 between the velocity-selection effect
 of the Bragg beams and the formation of a BEC-type state among some atoms.
 To solve this problem is a challenging task.
 In fact, the mechanism of the formation of BEC is a topic that has not been fully understood, yet,
 though lots of efforts have been seen and important progresses have been achieved
 \cite{Stoof99,Stoof01,Gardiner03,Gardiner08,Berloff,Blakie,Bezett,Kagan96,Kagan97,Castin,Calzetta,QK}.

 \acknowledgements

 W.W. is grateful to Jie Liu for initial stimulating discussions and is also grateful to Jiangbin
 Gong for valuable suggestions.
 This work was partially supported by the Natural Science Foundation of China
 under Grant Nos.~11275179 and 10975123 and the National Key Basic Research Program of China under Grant
 No.2013CB921800.

\begin{appendix}

\section{Derivation of Eq.(\ref{thermal_interference})} \label{app-thermal}

 In order to derive Eq.(\ref{thermal_interference}), let us first compute the phase difference between
 $\psi_1(t)$ and $\psi_2(t)$ in Eq.(\ref{psi-t}).
 At the time $t= \tau_f$, Eqs.(\ref{sigma}) and (\ref{con-xi}) give $\sigma \approx \frac{\hbar t}{2 m \xi}$.
 Then, making use of the expression in (\ref{sp}),
 it is easy to get the following expression of the phase difference,
 \begin{equation}
  -\frac{\vec{p}_0 \cdot \vec{d}}{\hbar}+\frac{[\vec{x}-(\vec{x_t}+\frac{\vec{d}}{2})]^2
 -[\vec{x}-(\vec{x_t}-\frac{\vec{d}}{2})]^2}{4 \sigma \xi},
 \end{equation}
 where
 \begin{equation}
 \vec{x}_t=\vec{x}_0+\frac{\vec{p}_0 t}{m}.
 \end{equation}
 Simple algebra shows that the difference has the following simple expression,
 \begin{equation} \label{pd}
  \frac{2 \pi (x-x_0)}{\lambda_f},
 \end{equation}
 where $\lambda_f$ is defined in Eq.(\ref{lf}).
 Then, the density is written as
\begin{equation}
\begin{split}
& \rho(\vec{x},\alpha,t) = |\psi(\vec{x},\alpha,t)|^2  \\
\approx &\frac{1}{2} A^6_\sigma e^{-\frac{(\vec{x}-\vec{x}_t)^2}{2 \sigma^2}} \left[e^{\frac{(x-x_t)d}{2\sigma^2}
}+e^{-\frac{(x-x_t)d}{2\sigma^2}}+2 \cos \frac{2 \pi (x-x_0)}{\lambda_f}\right]
\end{split}
 \label{s intf}
\end{equation}
 Integrating Eq.(\ref{s intf}) with the weight $f$ in Eq.(\ref{f-dis})
 over $\alpha=(\vec{x}_0, \vec{p}_0)$, one gets Eq.(\ref{thermal_interference}).

\section{Contrast in the simple thermal model with Bose-Einstein distribution} \label{sect-Bose}

\begin{figure} 
 \centering
 \includegraphics[width=\columnwidth]{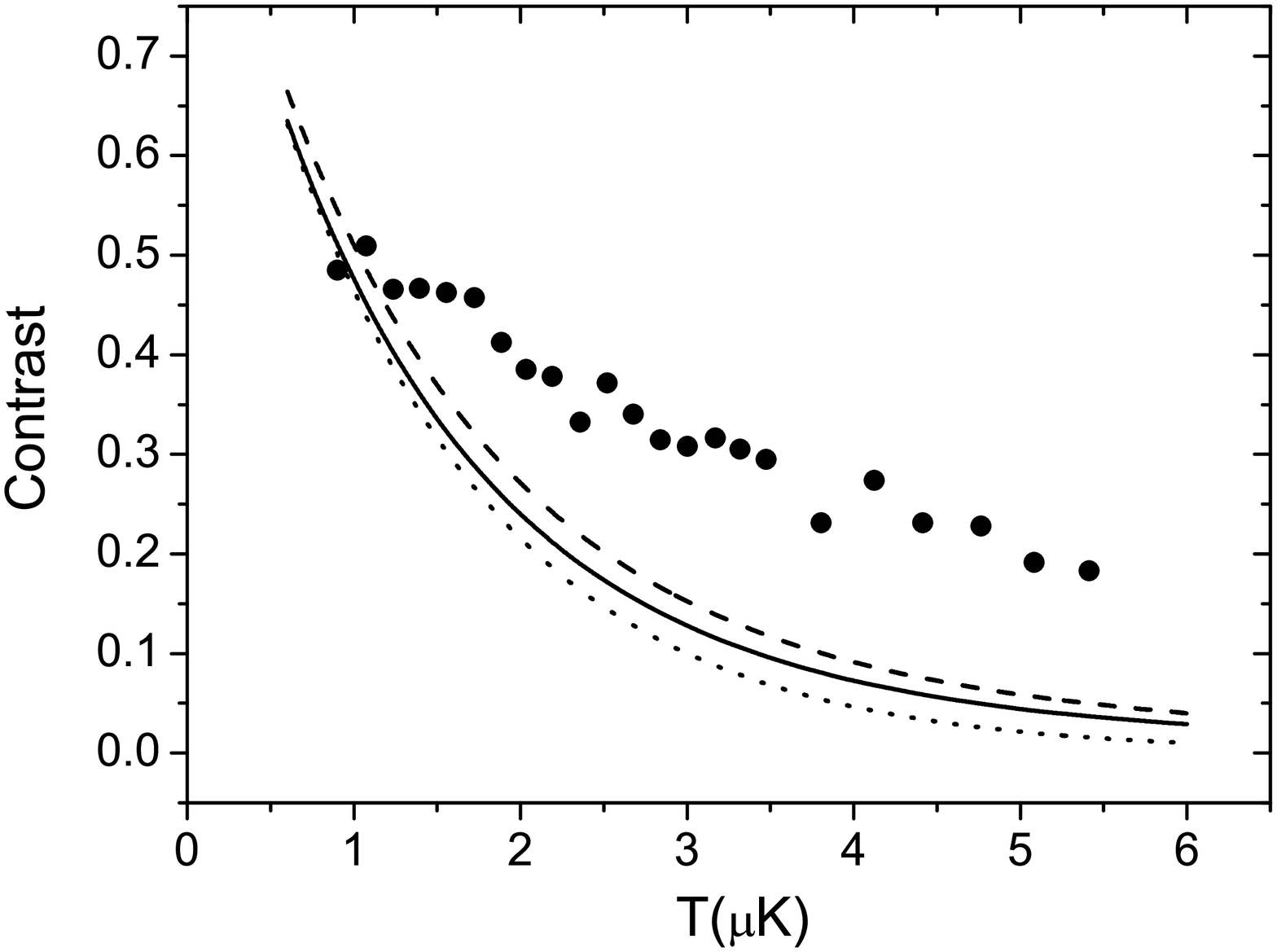}
 \vspace{0.4cm}
 \caption {Contrasts predicted by Eq.(\ref{COBE}), in the simple thermal model
 with the Bose-Einstein distribution.
  Dash curve: $\mu=0$. Solid curve: $\mu=-1 \mathrm{\mu K}$.
  For comparison, predictions of Eq.(\ref{Cth}) with the Boltzmann distribution (dotted curve)
  and the experimental data with $\lambda_f{=}170 \mathrm{\mu m}, \tau_p{=}30 \mathrm{\mu s}$ (solid
  circles) are also plotted.  }
 \label{comparison3}
 \end{figure}

 In this appendix, by numerical simulation, we show that, in the simple thermal model
 with the Bose-Einstein distribution for the function $f$ in Eq.(\ref{f-dis}),
 the obtained contrasts are still not close to the experimental results in the temperature region
 of $T>T_d$.
 The Bose-Einstein distribution is written as
\begin{equation}
f_{_\mathrm{BE}}(\vec{x_0},\vec{p_0})=\frac{1}{e^{(E-\mu)/k_B T}-1},
\label{BE}
\end{equation}
 where $E{=}\frac{\vec{p_0}^2}{2 m}{+}\frac{m \omega^2 \vec{x_0}^2}{2}$ is the
 single particle energy in the trap and $\mu$ is the chemical potential.
 Under this distribution, direct computation shows that, in the simple thermal model,
 the contrast has the following expression,
\begin{equation} \label{COBE}
 \ww C_{\rm th}=\frac{1}{\sum_{n=1}^{\infty}\frac{1}{n^3}z^n}
 \left \{ \sum_{n=1}^{\infty}\frac{1}{n^3}z^n \exp \left( -\frac{2 \pi^2 R^2_T}{n\lambda^2_f} \right) \right \},
\end{equation}
where $z{=}e^{\mu/k_BT}$.
 Taking the first-order terms in both the numerator and the denominator,
 Eq.(\ref{COBE}) gives $C_{\mathrm{th}}$ in Eq.(\ref{Cth}).
 For a Bose gas, $\mu{<}0$, hence $z{<}1$.
 Numerically, we found that predictions of Eq.(\ref{COBE}) are not close to
 the experimental data for $T>T_d$, as illustrated in Fig.\ref{comparison3}.

\section{ Contrast for a class of the initial condition} \label{un}

 In this appendix, we discuss the contrast under an initial condition of a type more generic
 than Gaussian wave packets.
 It is shown that the expression of the contrast $C_{\rm th}$ in Eq.(\ref{Cth}),
 as well as the expression of the fringe spacing $\lambda_f$ in Eq.(\ref{lf}) are still approximately
 valid for a more generic type of the single-particle states in the simple thermal model.

 For simplicity in discussion, we discuss within a 1-dimensional configuration space.
 We consider an initial packet $\varphi_0(x)$, centered at $x_0$ in the coordinate space
 and at $p_0$ in the momentum space.
 We assume that the main body of the packet lies in a region of a scale $l$,
 namely, $|x-x_0| \le l$, not necessarily of a Gaussian shape.
 Free expansion of the packet gives
\begin{equation}
  \varphi(x,t)=\frac{1}{2 \pi \hbar}\int \mathrm{d}p \mathrm{d}x' \exp\left[
  -\frac{i p^2 t}{2 m \hbar}+\frac{ip(x-x')}{\hbar}\right]\varphi_0(x').
 \end{equation}
 Changing the variable $p$ to $p'=p-\frac{m (x-x')}{t}$, then, integrating out $p'$, we get
\begin{equation}\label{varphi-t}
\varphi(x,t)=\sqrt{\frac{m}{2 i \pi \hbar  t} }\int \mathrm{d}x' \exp \left(\frac
{i (x-x')^2 m }{2 \hbar t}\right)\varphi_0(x').
\end{equation}

 Let us write the wave function beyond the two Bragg beams as
 \begin{equation}
 \psi_0(x)=\varphi_0(x)+\varphi_0(x+d).
 \end{equation}
 Making use of Eq.(\ref{varphi-t}), simple derivation gives the following expression
 for the time evolution of $\psi$,
 \begin{equation}\label{psi-phi0}
 \begin{split}
 &\psi(x,t)
 =\sqrt{\frac{m}{2i  \pi \hbar  t} }\int \mathrm{d}x' \exp
 \left( \frac {i (x-x')^2 m }{2 \hbar t}\right)\varphi_0(x') \\
 &\times \left[1+\exp\left(\frac{id(2x-2x'+d)m}{
 2 \hbar t}\right)\right].
 \end{split}
 \end{equation}
 For times sufficiently long, one has $l \ll \lambda $, where
\begin{equation}\label{}
 \lambda=\frac{h t}{m d}.
\end{equation}
 Since the main body of $\varphi_0(x')$ lies within a region of the scale $l$ centered at
 $x_0$, in the integration on the right hand side of Eq.(\ref{psi-phi0}),
 approximately, one may consider the integration domain $(x_0-l,x_0+l)$.
 Within this region, because of the relation $l \ll \lambda $,
 the variable $x'$ in the term $\exp\left(\frac{id(2x-2x'+d)m}{2 \hbar t}\right)$ can be approximately
 taken as $x_0$.
 Then, making use of Eq.(\ref{varphi-t}), Eq.(\ref{psi-phi0}) can be written as
\begin{equation}
\psi(x,t)\approx \varphi(x,t)\left[1+\exp \left(i\frac{2 \pi (x-x_0+\frac{d}{2})}{\lambda}\right)\right].
\end{equation}
 This gives
\begin{equation}\label{un_rho}
\rho(x,t) \approx |\varphi(x,t)|^2\left(1+\cos \frac{2 \pi (x-x_0+\frac{d}{2})}{\lambda}\right).
\end{equation}
 For a slowly-varying $|\varphi(x,t)|^2$, Eq.(\ref{un_rho}) predicts
 an interference pattern with a fringe spacing $\lambda$ under an envelope $|\varphi(x,t)|^2$.
 Note that $\lambda$ gives $\lambda_f$ in Eq.(\ref{lf}) at $t=\tau_f$.

 For an ensemble of the packets, with $x_0$ and $p_0$ obeying the Boltzmann distribution
 [cf.~Eq.(\ref{f-dis})],
 direct derivation gives the following expression for the density,
 \begin{equation}
 n(x) \approx F_0(x)+F_1(x)\cos\frac{2 \pi (x+\frac{d}{2})}{\lambda}
 +F_2(x)\sin\frac{2 \pi (x+\frac{d}{2})}{\lambda},
 \end{equation}
 where
\begin{eqnarray} \nonumber
 & F_0(x) &=\int \mathrm{d} x_0 G(x,x_0) \exp\left(-\frac{x_0^2}{2 R^2_T}\right),
 \\ & F_1(x) &=\int \mathrm{d} x_0 G(x,x_0) \exp\left(-\frac{x_0^2}{2 R^2_T}\right)
 \cos \frac{2 \pi x_0}{\lambda}, \nonumber
 \\ & F_2(x) & =\int \mathrm{d} x_0 G(x,x_0) \exp\left(-\frac{x_0^2}{2 R^2_T}\right)
 \sin \frac{2 \pi x_0}{\lambda}. \ \ \
\end{eqnarray}
 Here,
 \begin{equation}
 G(x,x_0)=\frac{\hbar \omega}{k_B T} \int \mathrm{d} p_0 |\varphi(x,t;x_0,p_0)|^2 \exp\left(-\frac{p^2_0}{2 m k_B T}\right),
 \end{equation}
 with the dependence on $x_0$ and $p_0$ written explicitly.

 In some situations of interest, the quantity $G(x,x_0)$ can be approximately regarded as a constant
 for $x$ in the region of measurement, i.e., for $x$ of the order of $\lambda$,
 and for $x_0$ in a region wherein $\exp\left(-\frac{x_0^2}{2 R^2_T}\right)$ is not small.
 To be specific, we discuss two examples.
 In the first example, $|\varphi|^2 \propto \delta(x-x_0-p_0 t/m)$.
 This gives
 \begin{equation}
 G(x,x_0) \propto \exp\left[-\frac{(x-x_0)^2}{2 L^2_t}\right],
 \end{equation}
 where $L_t=\sqrt{k_B T t^2 /m}$.
 For the parameters used in the experiments in Ref.\cite{Miller}, direct computation shows that
 $\lambda /L_t =(\sqrt{2 \pi} \lambda_T)/d$ is approximately between $1/3$ and $1/15$,
 and $R_T/L_t = 1/\omega t \approx 1/20$ for $t=\tau_f$.
 Hence, $G$ is approximately a constant for the value of $x$ and $x_0$ of interest here.
 In the second example, $|\varphi|^2 $ has an Gaussian form with a standard deviation  $\sigma$, i.e.,
 $|\varphi|^2 \propto \exp[-(x-x_0-p_0 t/m)^2/(2 \sigma^2)]$, like that in the thermal model discussed above.
 In this case,
 \begin{equation}
 G(x,x_0) \propto \exp\left[-\frac{(x-x_0)^2}{2 (L^2_t+\sigma^2)}\right].
 \end{equation}
 Similarly, $G$ is approximately a constant in the region of interest here.

 For an approximately constant $G$, $F_2 \approx 0$ and can be neglected.
 Then, direct computation shows that
 \begin{equation}
 n(x) \approx F_0\left[1+C_{\rm th} \cos \frac{2 \pi (x+\frac{d}{2})}{\lambda}\right],
 \end{equation}
 where $C_{\rm th}$ is the contrast given in the simple thermal model in Eq.(\ref{Cth}).

\end{appendix}


\begin{thebibliography}{01}
\bibitem{footnote-c1}
 With further increase of the temperature $T$, the body of the atoms in mutual-coherence may split,
 then, the value of $c_1$ may decrease considerably.
\end{thebibliography}
 \end{document}